\documentclass[usegraphicx,usenatbib]{mn2e}

\begin{document}
\title[Primordial Non--Gaussianity and Large-Scale
Structure]{The Effect of Primordial Non--Gaussianity on the Topology
of Large-Scale Structure}
\author[Hikage et al.]{
C.~Hikage$^{1,2}$\thanks{E-mail: chiaki.hikage@nottingham.ac.uk},
P.~Coles$^2$, M.~Grossi$^{3}$, L.~Moscardini$^{4,5}$, K.~Dolag$^3$,
E.~Branchini$^6$,
\newauthor S.~Matarrese$^{7,8}$ \\
$^1$
School of Physics and Astronomy, University of Nottingham,
University Park, Nottingham, NG7 2RD \\
$^2$
 School of Physics and Astronomy, Cardiff University,
 5, The Parade, Cardiff, CF24 3AA \\
$^3$
 Max-Planck Institut f\"ur Astrophysik, Karl-Schwarzschild Strasse 1,
 D-85748 Garching, Germany \\
$^4$
 Dipartimento di Astronomia, Universit$\grave{\rm a}$ di Bologna,
 via Ranzani 1, I-40127 Bologna, Italy \\
$^5$
 INFN, Sezione di Bologna, viale Berti Pichat 6/2, I-40127 Bologna, Italy \\
$^6$
 Dipartimento di Fisica, Universit$\grave{\rm a}$ di Roma TRE,
 via della Vasca Navale 84, I-40127 Bologna, Italy \\
$^7$
 Dipartimento di Fisica, Universit$\grave{\rm a}$ degli Studi di Padova,
 via Marzolo 8, I-35131, Padova, Italy \\
$^8$
 INFN, Sezione di Padova, via Marzolo 8, I-35131, Padova, Italy
}
%\date{\today}
\date{Accepted 2008 January 10. Submitted 2007 November 19.}
\maketitle
%%%%%%%%%%%%%%%%%%%%%%%%%%%%%%%%%%%%%%%%%%%%%%%%%%%%%%%
\begin{abstract}
We study the effect of primordial non--Gaussianity on the development
of large-scale cosmic structure using high-resolution $N$-body
simulations. In particular, we focus on the topological properties of
the ``cosmic web'', quantitatively characterized by the Minkowski
Functionals, for models with quadratic non-linearities with different
values of the usual non--Gaussianity parameter $f_{\rm NL}$. In the
weakly non-linear regime (the amplitude of mass density fluctuations
$\sigma_0<0.1$), we find that analytic formulae derived from
perturbation theory agree with the numerical results within a few
percent of the amplitude of each MF when $|f_{\rm NL}|<1000$.  In the
non-linear regime, the detailed behavior of the MFs as functions of
threshold density deviates more strongly from the analytical curves,
while the overall amplitude of the primordial non--Gaussian effect
remains comparable to the perturbative prediction. When smaller-scale
information is included, the influence of primordial non--Gaussianity
becomes increasingly significant statistically due to decreasing
sample variance. We find that the effect of the primordial
non-Gaussianity with $|f_{\rm NL}|=50$ is comparable to the sample
variance of mass density fields with a volume of $0.125(h^{-1}{\rm
Gpc})^3$ when they are smoothed by Gaussian filter at a scale of
$5h^{-1}$Mpc. The detectability of this effect in actual galaxy
surveys will strongly depend upon residual uncertainties in
cosmological parameters and galaxy biasing.
\end{abstract}
%%%%%%%%%%%%%%%%%%%%%%%%%%%%%%%%%%%%%%%%%%%%%%%%%%%%%%%
%%%%%%%%%%%%%%%%%%%%%%%%%%%%%%%%%%%%%%%%%%%%%%%%%%%%%%%

\section{Introduction}

According to the standard scenarios for the formation of
large-structure in the Universe, the present-day cosmic density
field evolves from small-amplitude initial fluctuations which are
described by Gaussian statistics. The hypothesis of primordial
Gaussianity is  supported by present observations of the Cosmic
Microwave Background (CMB), particularly those from the Wilkinson
Microwave Anisotropy Probe (WMAP) \citep{Komatsu2003,Spergel2007}.
These results are consistent with an inflationary origin
for the primordial perturbations, since the simplest forms of cosmic
inflation produce nearly Gaussian fluctuations.

In order to understand the early Universe in more detail, however, it
is necessary to measure (or at least constrain) the departures from
non--Gaussianity that inevitably arise at some level during the
inflationary epoch. For example, the simplest slowly-rolling single
field inflation model predicts very small levels of primordial
non--Gaussianity, while multi-field inflation models and models with a
non-standard kinetic term for the inflation may yield larger effects
which could be detected in ongoing or next-generation observations
\citep[e.g.][]{BMR2002,BU2002,lyth2003,dvali2004,ACMZ2004,AST2004,BKMR2004,Chen2007,BB2007}.
Only when such phenomena are detected will it be possible to
distinguish between the hundreds of currently viable variations on the
theme of inflation by understanding the dynamical behavior of the
inflation field.

In order to model the primordial non--Gaussianity that might arise
during inflation, the following simple form including quadratic
corrections to the curvature perturbation $\Phi$ \citep{B1980} during
the matter era has been often adopted
\citep{Gangui1994,Verde2000,KS2001}:
%%%%%%%%%%%%%%%%%%%%%%%%%%%%%%%%%%%%%%%%%%%%%%%%%%%%%%%
\begin{equation}
\label{eq:ngpotential2}
\Phi=\phi+f_{\rm NL}(\phi^2-\langle\phi^2\rangle),
\end{equation}
%%%%%%%%%%%%%%%%%%%%%%%%%%%%%%%%%%%%%%%%%%%%%%%%%%%%%%%
where $\phi$ represents an auxiliary random-Gaussian field and $f_{\rm
NL}$ characterizes the amplitude of a quadratic correction to the
curvature perturbations in a dimensionless way. In principle, $f_{\rm
NL}$ could be scale-dependent, but current observations are not
sufficiently sensitive to detect any such variation, so a constant
$f_{\rm NL}$ remains a useful parametrization of the level of
non--Gaussianity. Recent analyses of the angular bispectrum for WMAP
provides strong constraints on $f_{\rm NL}$ to lie in the range from
$-54$ to $114$ at the 95 percent confidence level
\citep{Komatsu2003,Spergel2007,Creminelli2006}.

The Large-Scale Structure (LSS) of the distribution of galaxies in the
Universe provides another potentially powerful probe of primordial
non--Gaussianity
\citep{FS1994,CB1996,Verde2000,Scocci2004,HKM2006,SK2007}.  The
three-dimensional spatial information arising from LSS is potentially
a richer source information about primordial non--Gaussianity than the
two-dimensional information arising from the CMB.  For example,
constraints from upcoming cluster surveys should be comparable with
current CMB limits and those from galaxy surveys, which could be as
tight as $|f_{\rm NL}|\sim 10$ for the planned surveys and $|f_{\rm
NL}|\sim 0.2$ for an all-sky survey of galaxies up to redshift $z=5$
\citep{SK2007,Dalal2007}.  A variety of large-scale projects of LSS
observation covering Gpc$^3$ volumes are being proposed, such as an
extension of the Sloan Digital Sky Survey; APO-LSS survey; The
Hobby-Eberly Dark Energy Experiment (HETDEX) \citep{Hill2004};
Wide-Field Multi-Object Spectrograph (WFMOS) \citep{Glazebrook2005};
and the Cosmic Inflation Probe (CIP) mission \citep{Melnick2004}.  It
is consequently important to study the optimal way to extract
information about primordial non--Gaussianity from such surveys.

The statistical analysis of non--Gaussianity has been mainly performed
through the calculation of the bispectrum
\citep{Verde2000,Scocci2004,SK2007}.  Strong motivation for this is
that the bispectrum is the simplest statistical function that can
measure quadratic non-linearity \citep[e.g.][]{WC2003}. Although
the quadratic model provides an extremely useful benchmark for
statistical analysis techniques, one must always bear in mind that
there are many different ways for a random field to be
non--Gaussian. In general, there is no one statistic that completely
characterizes the statistical nature of a non--Gaussian random field,
so a battery of higher-order statistics must be deployed. In
particular, when the full nature of non--Gaussianity is virtually
unknown, such as is really the case for primordial perturbations, the
theoretical model assumed should be validated before its parameters
are constrained. Different statistics reflect different aspects of
non--Gaussianity so the use of different statistics plays a vital role
in this kind of consistency check.

In this paper we use a set of invariant characteristics of the
topology of the cosmic web, known as the Minkowski Functionals
(MFs). These have already been used to describe the morphological
properties of cosmic density fields in a variety of contexts
\citep{MBW1994,SB1997,SG1998,Hikage2003}. Four MFs are defined in
three-dimensional density fields such as LSS: the volume
fraction ($V_0$); surface area ($V_1$); mean curvature ($V_2$);
and Euler characteristic ($V_3$).

Using a perturbative approach, \citet{HKM2006} derived analytical
formulae for the behavior of the MFs for LSS including primordial
non--Gaussianity (as a function of $f_{\rm NL}$ as given in equation
[\ref{eq:ngpotential2}]), in addition to the non--Gaussianity due to
non-linear gravity and galaxy biasing. The validity of the
perturbative analysis is, however, limited to the weakly non-linear
regime.  Smaller-scale modes also contain rich information about the
primordial density fields, and this could help place more stringent
constraints on primordial non--Gaussianity. In this paper, we use
high-resolution $N$-body simulations to study the effect of primordial
non--Gaussianity on the MFs from the mildly to strongly non-linear
regime. There are two reasons for using the full numerical analysis:
one is to see how well the perturbative formulae describe the
simulated MFs to check their applicability; the other is to study how
the primordial non--Gaussian effect behaves in the strongly non-linear
regime and thus to estimate the significance of the effect on the MFs.

The paper is organized as follows. In Section \ref{sec:pb}, we
review the perturbative formulae for the MFs.  The details of the
$N$-body simulations and the computing method of the MFs are
summarized in Section \ref{sec:sim}. In Section \ref{sec:results},
we compare the perturbative formulae of MFs with simulated results to
study the primordial non--Gaussian effect in non-linear regime.
Section \ref{sec:summary} is devoted to the summary and conclusions.

%%%%%%%%%%%%%%%%%%%%%%%%%%%%%%%%%%%%%%%%%%%%%%%%%%%%%%%
\section{Perturbation Theory}
\label{sec:pb}

We define the MFs of density fields for a given threshold $\nu\equiv
\delta/\sigma_0$, where $\delta$ is the density fluctuation, which
has zero mean, and $\sigma_0\equiv \langle\delta^2\rangle^{1/2}$ is
its standard deviation. The $k$-th MF $V_k(\nu)$ can be written
separately with the amplitude $A_k$ and the function of $\nu$,
$v_k(\nu)$, as
%%%%%%%%%%%%%%%%%%%%%%%%%%%%%%%%%%%%%%%%%%%%%%%%%%%%%%%%%%%%%%%%%%%%
\begin{equation}
\label{eq:MFs}
V_k(\nu) = A_kv_k(\nu).
\end{equation}
%%%%%%%%%%%%%%%%%%%%%%%%%%%%%%%%%%%%%%%%%%%%%%%%%%%%%%%%%%%%%%%%%%%%
The amplitude part $A_k$, which depends only on the power spectrum $P(k,z)$
of the 3-dimensional fluctuation field $\delta$ at redshift $z$, is given by
%%%%%%%%%%%%%%%%%%%%%%%%%%%%%%%%%%%%%%%%%%%%%%%%%%%%%%%%%%%%%%%%%%%%
\begin{equation}
\label{eq:mfamp}
A_k = \frac1{(2\pi)^{(k+1)/2}}\frac{\omega_3}{\omega_{3-k}\omega_k}
\left(\frac{\sigma_1(z)}{\sqrt{3}\sigma_0(z)}\right)^k,
\end{equation}
%%%%%%%%%%%%%%%%%%%%%%%%%%%%%%%%%%%%%%%%%%%%%%%%%%%%%%%%%%%%%%%%%%%%
where $\omega_k\equiv\pi^{k/2}/{\Gamma(k/2+1)}$ gives $\omega_0=1$,
$\omega_1=2$, $\omega_2=\pi$, and $\omega_3=4\pi/3$. The quantity
$\sigma_i^2$ characterizes the variance of fluctuating fields for
$i=0$ and that of their derivatives for $i=1$ given by
%%%%%%%%%%%%%%%%%%%%%%%%%%%%%%%%%%%%%%%%%%%%%%%%%%%%%%%%%%%%%%%%%%%%
\begin{equation}
\sigma_i^2(z)\equiv \int_0^\infty\frac{k^2dk}{2\pi^2}k^{2i}P(k,z)W^2(kR),
\end{equation}
%%%%%%%%%%%%%%%%%%%%%%%%%%%%%%%%%%%%%%%%%%%%%%%%%%%%%%%%%%%%%%%%%%%%
where $W$ represents a smoothing kernel. Throughout the paper, we
adopt a Gaussian kernel $W^2=\exp[-(kR)^2]$ where $R$ represents the
smoothing scale.

\citet{Matsubara2003} derives the second-order perturbative formulae
of the MFs using the multivariate Edgeworth expansion.  According
to the formulae, the function $v_k(\nu)$ is written with the Gaussian
part $v_k^{\rm (G)}$ and the leading part of the non--Gaussian term
$\Delta v_k$ as
%%%%%%%%%%%%%%%%%%%%%%%%%%%%%%%%%%%%%%%%%%%%%%%%%%%%%%%%%%%%%%%%%%%%
\begin{eqnarray}
v_k(\nu) & = & v_k^{\rm (G)}(\nu) + \Delta v_k(\nu), \\
v_k^{\rm (G)}(\nu) & = & e^{-\nu^2/2}H_{k-1}(\nu), \\
\label{eq:MFs_perturb}
\Delta v_k(\nu) & = & e^{-\nu^2/2}
\left[\frac16S^{(0)}H_{k+2}(\nu)
+\frac{k}3S^{(1)}H_k(\nu)\right. \nonumber \\
&+&\left.\frac{k(k-1)}6S^{(2)}H_{k-2}(\nu)\right]\sigma_0,
\end{eqnarray}
%%%%%%%%%%%%%%%%%%%%%%%%%%%%%%%%%%%%%%%%%%%%%%%%%%%%%%%%%%%%%%%%%%%%
where $H_n(\nu)$ denote the Hermite polynomials.  The leading-order
non--Gaussian term $\Delta v_k(\nu)$ is calculated when the three
``skewness parameters'' $S^{(i)}$ are given.

The three skewness parameters $S^{(i)} (i=0,1$ and 2) are computed by
integrating the bispectrum $B(k_1,k_2,k_3,z)$ over $k_1$,
$k_2$, and $\mu\equiv({\mathbf k_1}\cdot{\mathbf k_2})/(k_1k_2)$ with
appropriate weights as
\citep{HKM2006}
%%%%%%%%%%%%%%%%%%%%%%%%%%%%%%%%%%%%%%%%%%%%%%%%%%%%%%%
\begin{eqnarray}
\label{eq:s0_lss}
S^{(0)}(z)&=&\frac{1}{8\pi^4\sigma_0^4(z)}
\int^\infty_0 dk_1 \int^\infty_0 dk_2 \int^1_{-1} d\mu
k_1^2k_2^2 \\ \nonumber
& &B(k_1,k_2,k_{12},z)W(k_1R)W(k_2R)W(k_{12}R),\\
\nonumber
S^{(1)}(z)&=&
\frac{1}{16\pi^4\sigma_0^2(z)\sigma_1^2(z)}
\int^\infty_0 dk_1 \int^\infty_0 dk_2 \int^1_{-1} d\mu \\ \nonumber
& &k_1^2k_2^2(k_1^2+k_2^2+\mu k_1k_2)
B(k_1,k_2,k_{12},z)\\
\label{eq:s1_lss}
& &\times W(k_1R)W(k_2R)W(k_{12}R),\\
\nonumber
S^{(2)}(z)&=&\frac{3}{16\pi^4\sigma_1^4(z)}
\int^\infty_0 dk_1 \int^\infty_0 dk_2 \int^1_{-1} d\mu \\ \nonumber
& &k_1^4k_2^4(1-\mu^2)B(k_1,k_2,k_{12},z)\\
\label{eq:s2_lss}
& &\times W(k_1R)W(k_2R)W(k_{12}R),
\end{eqnarray}
%%%%%%%%%%%%%%%%%%%%%%%%%%%%%%%%%%%%%%%%%%%%%%%%%%%%%%%
where $k_{12}\equiv|{\mathbf k_1}+{\mathbf k_2}|=(k_1^2 +k_2^2+2\mu
k_1k_2)^{1/2}$.

Throughout this paper, we neglect the non--Gaussianity arising from
the non-linearity in relationship between galaxy counts
and mass (i.e. galaxy biasing)
so as to keep the analysis as simple as possible.
The bispectrum $B$ for the matter density fluctuation is then given by
%%%%%%%%%%%%%%%%%%%%%%%%%%%%%%%%%%%%%%%%%%%%%%%%%%%%%%%
\begin{equation}
B(k_1,k_2,k_3,z)= B_{\rm pri}(k_1,k_2,k_3,z)
+B_{\rm gr}(k_1,k_2,k_3,z),
\end{equation}
%%%%%%%%%%%%%%%%%%%%%%%%%%%%%%%%%%%%%%%%%%%%%%%%%%%%%%%
where $B_{\rm pri}$ and $B_{\rm gr}$ represent the
contributions from primordial non--Gaussianity and non-linearity in
gravitational clustering respectively:
%%%%%%%%%%%%%%%%%%%%%%%%%%%%%%%%%%%%%%%%%%%%%%%%%%%%%%%
\begin{eqnarray}
 B_{\rm pri}(k_1,k_2,k_3,z)
&\equiv&
 \frac{2f_{\rm NL}}{D(z)}
\left[\frac{P(k_1,z)P(k_2,z)M(k_3)}{M(k_1)M(k_2)}\right.
\nonumber \\
&+&\left.(\mbox{cyc.})\right],\\
B_{\rm gr}(k_1,k_2,k_3,z)
&\equiv&
 2\left[F_2({\mathbf k}_1,{\mathbf k}_2)P(k_1,z)
P(k_2,z)\right. \nonumber \\
&+&\left.(\mbox{cyc.})\right],
\end{eqnarray}
%%%%%%%%%%%%%%%%%%%%%%%%%%%%%%%%%%%%%%%%%%%%%%%%%%%%%%%
where $D(z)$ is the growth rate of linear density fluctuations
normalized such that $D(z)\rightarrow 1/(1+z)$ during the matter
era. The function $M(k)$ and $F_2({\mathbf k}_1,{\mathbf k}_2)$ are
 time-independent kernels describing mode-coupling due to
non-linear clustering of matter density fluctuations in the weakly
non-linear regime. These are given by
%%%%%%%%%%%%%%%%%%%%%%%%%%%%%%%%%%%%%%%%%%%%%%%%%%%%%%%
\begin{equation}
 M(k)\equiv \frac23\frac{k^2T(k)}{\Omega_{\rm m}H_0^2},
\label{eq:Mk}
\end{equation}
%%%%%%%%%%%%%%%%%%%%%%%%%%%%%%%%%%%%%%%%%%%%%%%%%%%%%%%
\begin{equation}
\label{eq:f2}
F_2({\mathbf k}_1,{\mathbf k}_2)=\frac{5}{7}
+\frac{{\mathbf k}_1\cdot{\mathbf k}_2}{
2k_1k_2}\left(\frac{k_1}{k_2}+\frac{k_2}{k_1}\right)+\frac{2}{7}
\frac{({\mathbf k}_1\cdot{\mathbf k}_2)^2}{k_1^2k_2^2}.
\end{equation}
%%%%%%%%%%%%%%%%%%%%%%%%%%%%%%%%%%%%%%%%%%%%%%%%%%%%%%%
We adopt the linear transfer function $T(k)$ by \citet{EH1999}.
In comparison with numerical simulations, we use the power spectrum of
the simulations (the details are explained in the next section) at
$z^\ast=76.97$ for a theoretical input of the power spectrum
$P(k,z^\ast)$ and then give the power spectrum at $z$ as
%%%%%%%%%%%%%%%%%%%%%%%%%%%%%%%%%%%%%%%%%%%%%%%%%%%%%%%
\begin{equation}
P(k,z)=\frac{D^2(z)}{D^2(z^\ast)}P(k,z^\ast).
\end{equation}
%%%%%%%%%%%%%%%%%%%%%%%%%%%%%%%%%%%%%%%%%%%%%%%%%%%%%%%

%%%%%%%%%%%%%%%%%%%%%%%%%%%%%%%%%%%%%%%%%%%%%%%%%%%%%%%
\section{Methodology}
\label{sec:sim}
\subsection{Numerical Simulations with Primordial Non--Gaussianity}

The $N$-body simulations with primordial non--Gaussianity that we use
for this analysis are those described in \citet{Grossi2007}.  These
simulations employ $800^3$ dark matter particles in a periodic cubic
box with a side length of 0.5$h^{-1}$Gpc. The cosmology of our
simulations is a flat $\Lambda$CDM model with mass density parameter
$\Omega_{\rm m}=0.3$, baryon density parameter $\Omega_{\rm b}=0.04$,
Hubble parameter $h=0.7$, primordial power-law index $n_{\rm s}=1$, and
$\sigma_8=0.9$.

The initial particles are perturbed from an initially homogeneous
``glass-like'' distribution. The primordial non--Gaussianity is
incorporated into a Gaussian-random field with the above cosmology
in the form of equation (\ref{eq:ngpotential2}). \citet{Grossi2007}
explored 7 different scenarios with $f_{\rm NL}=0,\pm 100, \pm 500 \
{\rm and} \pm 1000$. We have analyzed all of these simulations, but
for brevity in this paper we only present results for the Gaussian
simulation with $f_{\rm NL}=0$ and the two extreme non--Gaussian
cases $f_{\rm NL}=\pm 1000$; results for the other simulations with
$f_{\rm NL}=\pm 100$ are intermediate, as expected.

After Fourier-transforming the primordial non--Gaussian field, the
dark matter particles are displaced on the initial grid assuming the
Zel'dovich approximation.  The simulations are started at $z\approx
100$ and the subsequent gravitational evolution is simulated with the
GADGET-2 code \citep{Springel2005}. The Triangular-Shaped Cloud method
is used to assign densities onto $512^3$ grids. After
Fourier-transforming the grid data, we multiply by the Gaussian kernel
$\exp[-(kR)^2]$, and then transform them back to real space.

%%%%%%%%%%%%%%%%%%%%%%%%%%%%%%%%%%%%%%%%%%%%%%%%%%%%%%%
\begin{figure*}
\begin{center}
\includegraphics[width=18cm]{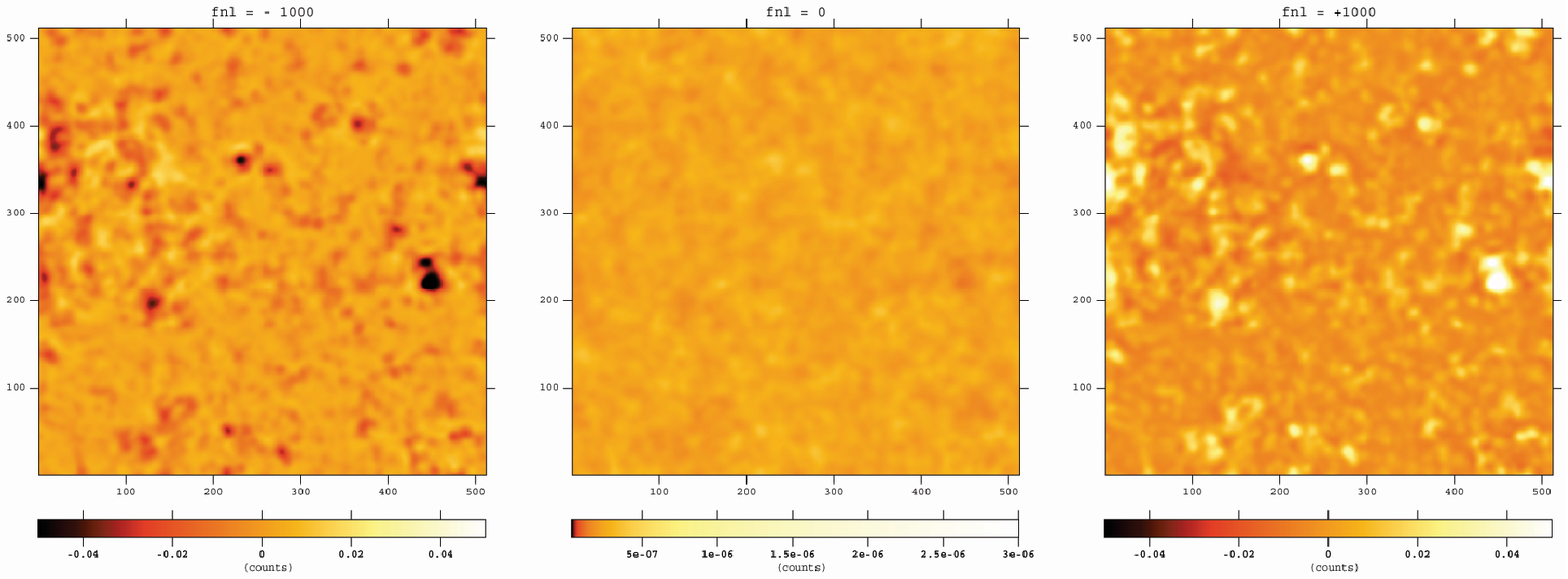}
\includegraphics[width=18cm]{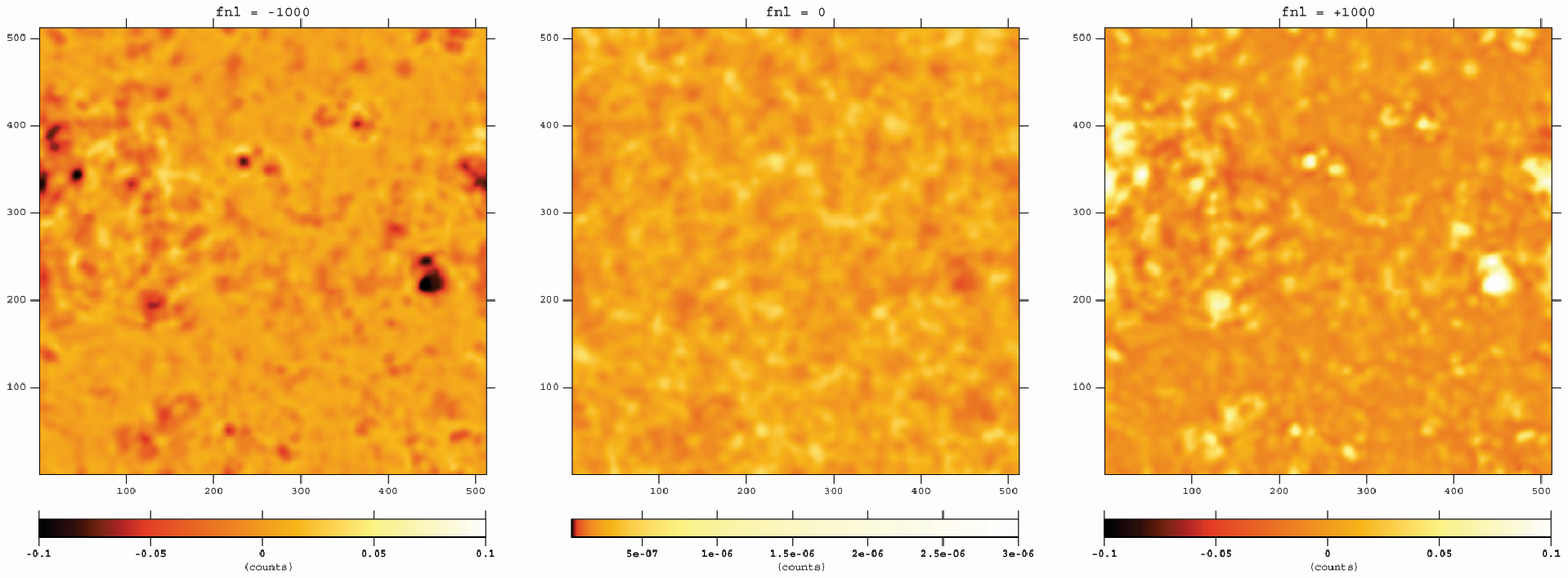}
\includegraphics[width=18cm]{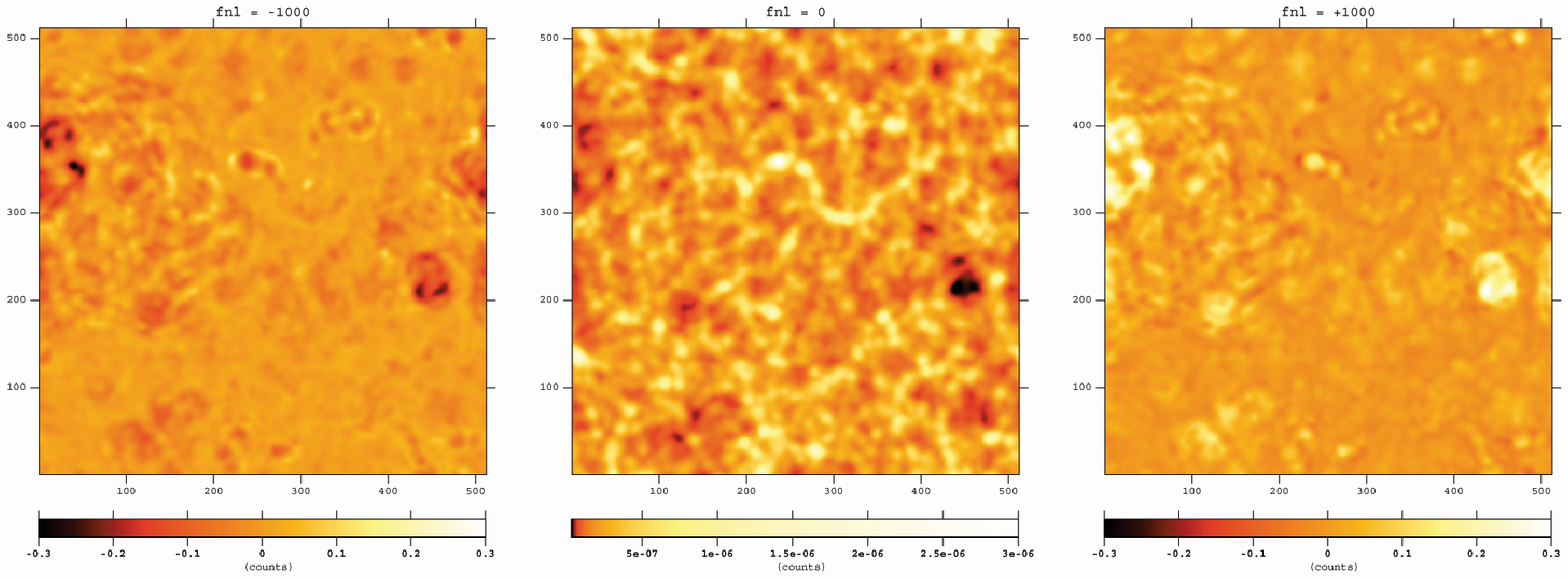}
\caption{Slice maps of simulated mass density fields at $z=5.15$
({\it top}), $z=2.13$ ({\it middle}) and $z=0$ ({\it bottom}). The
number of pixels at a side length is $512$ ($500h^{-1}$Mpc) and that
of the thickness is $32$ ($31.25h^{-1}$Mpc). The panels in the
middle row show the log of the projected density smoothed with a
Gaussian filter of 10 pixels width, corresponding to $9.8h^{-1}$Mpc.
The left and right panels are the relative residuals for the $f_{\rm
NL}$=$\pm 1000$ runs (equation [\ref{eq:nresidual}]). Each panel
has the corresponding color bar and the range considered are
different from panel to panel.}
\label{fig:1}
\end{center}
\end{figure*}
%%%%%%%%%%%%%%%%%%%%%%%%%%%%%%%%%%%%%%%%%%%%%%%%%%%%%%%

It is instructive first to examine the visual morphology of the
clustering pattern. Fig. \ref{fig:1} shows maps of slices of the
mass density field with $f_{\rm NL}=0$ (middle-row panels) and the
relative residuals between $f_{\rm NL}=\pm 1000$ and $f_{\rm NL}=0$
(left and right panels). The residual for the map with $f_{\rm
NL}=x$, $\Delta \rho_x$, is calculated at each pixel as
%%%%%%%%%%%%%%%%%%%%%%%%%%%%%%%%%%%%%%%%%%%%%%%%%%%%%%%
\begin{equation}
\Delta \rho_x = (\rho_x-\rho_0)/\rho_0
\label{eq:nresidual}
\end{equation}
%%%%%%%%%%%%%%%%%%%%%%%%%%%%%%%%%%%%%%%%%%%%%%%%%%%%%%%
where $\rho_x$ is the number density of mass particles for the map
with $f_{\rm NL}=x$. The field is smoothed with a Gaussian filter
10 pixels  wide (i.e. $9.8h^{-1}$Mpc). The redshifts of the maps
are $5.15, 2.13$ and $0$ from top to bottom respectively. Similar
density structures in the mass distribution appear in the residual
maps with their contrast at same (inverse) sign for positive
(negative) values of $f_{\rm NL}$. For example, a large void
structure at the right-center in the density map also appears in the
residual maps. This is because the higher density region is
initially more (less) enhanced in the positive (negative) $f_{\rm
NL}$, as predicted by the local model of primordial
non--Gaussianity in equation (\ref{eq:ngpotential2}).

\subsection{Computation of Minkowski Functionals}

The computational method we use for calculating MFs of data defined on
a grid is based on ideas from integral geometry, rather than the
alternative more cumbersome approach of using the differential
properties of bounding surfaces. In our case the calculation reduces
to counting the numbers of vertices, edges and sides of the elementary
cells covering the structure \citep{CDP1996,SB1997}. The range of
$\nu$ is from $-3.6$ to $3.6$ with an equal binning width of
$0.2$. The MFs measured from numerical simulations often deviate from
analytical predictions even for Gaussian realizations due to subtle
pixelization effects.  However, as pointed out by \citet{HKM2006},
pixelization effects become negligible when computing the difference
between Gaussian and non--Gaussian MFs. Therefore we focus on $\Delta
v_k(\nu_i)$ ($i$ denoting the binning number of $\nu$) that we compute
as follows:
\begin{enumerate}

\item  We compute the MFs for non--Gaussian simulation data $V_k$ and
 then divide them by their amplitudes $A_k$ (equation
 [\ref{eq:mfamp}]) to obtain normalized MFs $v_k$.  The $\sigma_0$ and
 $\sigma_1$ in $A_k$ are computed from the density fields of the
 simulations. \\

\item   The MFs for Gaussian fields are computed
 in the same way and then divided by their amplitudes $A_k$ where the
 values of $\sigma_0$ and $\sigma_1$ are computed from each
 realization.  The same cosmological parameters as the $N$-body
 simulations are adopted. The normalized MFs $v_k^{\rm (G)}$ are estimated
 by averaging MFs over 10 Gaussian realizations. \\

\item  The difference ratio  $\Delta v_k$ is computed by
\begin{equation}
 \Delta v_k=v_k-v_k^{\rm (G)}.
\end{equation}

\end{enumerate}

%%%%%%%%%%%%%%%%%%%%%%%%%%%%%%%%%%%%%%%%%%%%%%%%%%%%%%%
\section{Results}
\label{sec:results}

In this section we explore two different but related issues. The
first is whether the non-linear behaviour seen in numerical
simulations matches the predictions of analytical approaches. The
second is whether it is possible to separate the effects of
non-linear evolution from primordial non-Gaussianity to a sufficient
extent for this method to be useful in practice.
\subsection{Agreement with perturbative formulae
in the weakly non-linear regime}

%%%%%%%%%%%%%%%%%%%%%%%%%%%%%%%%%%%%%%%%%%%%%%%%%%%%%%%
\begin{figure*}
\begin{center}
\includegraphics[width=8cm]{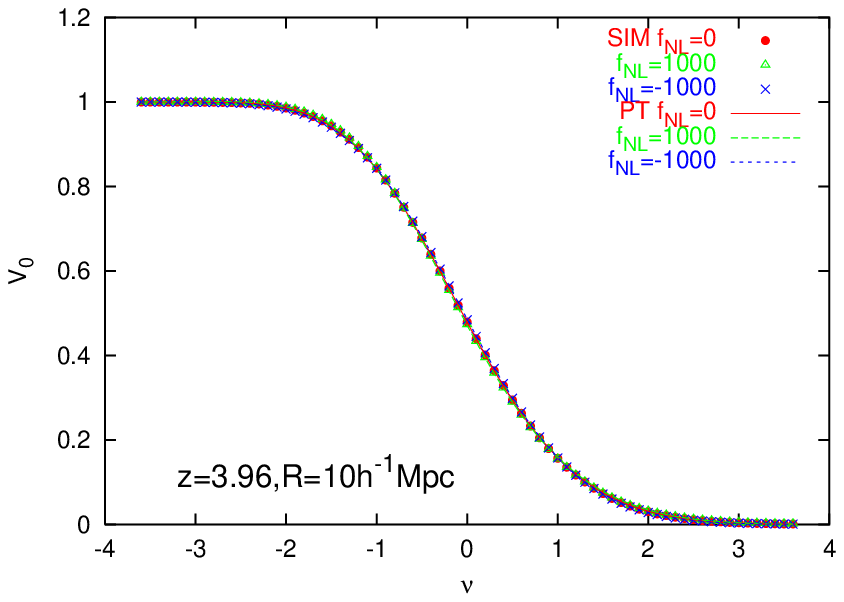}
\includegraphics[width=8cm]{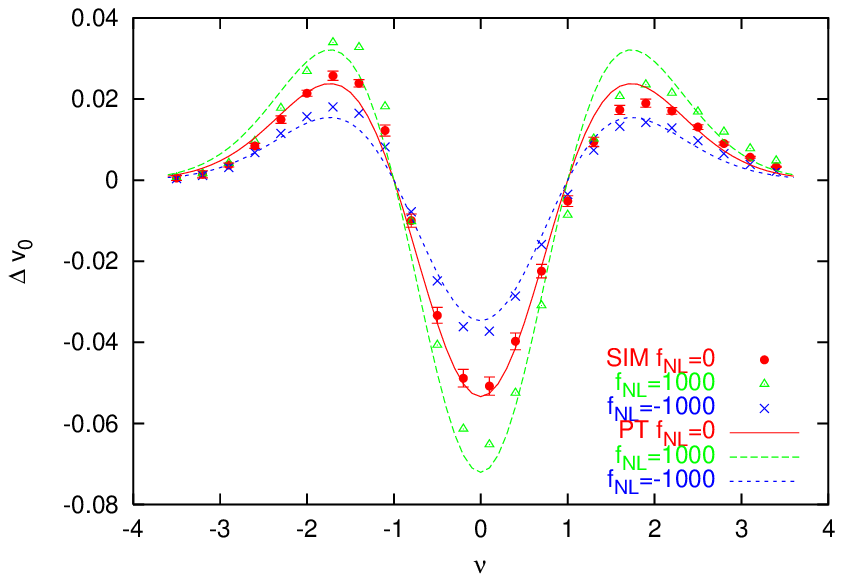}
\includegraphics[width=8cm]{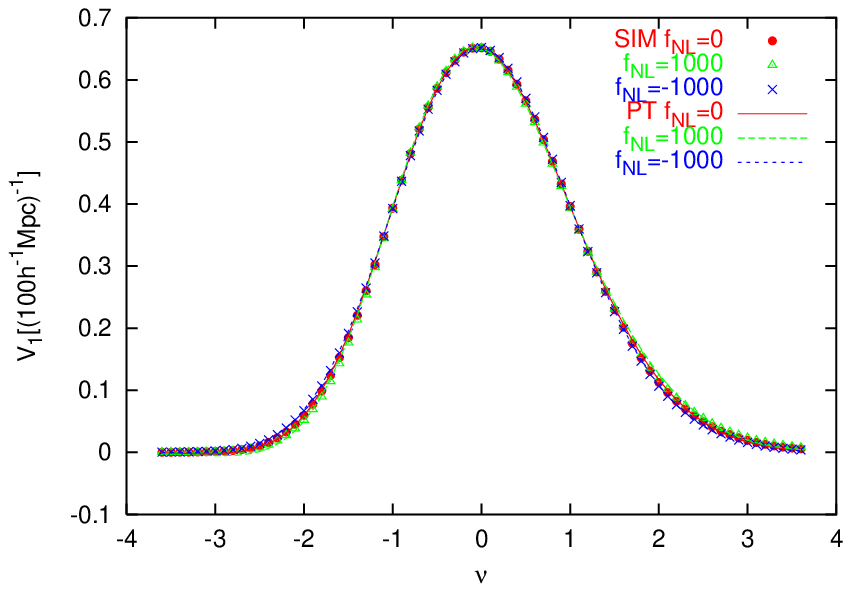}
\includegraphics[width=8cm]{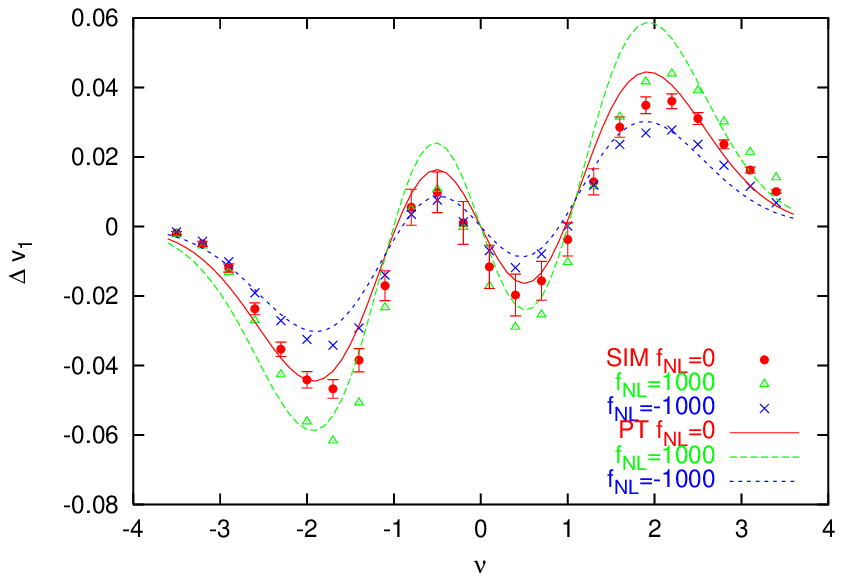}
\includegraphics[width=8cm]{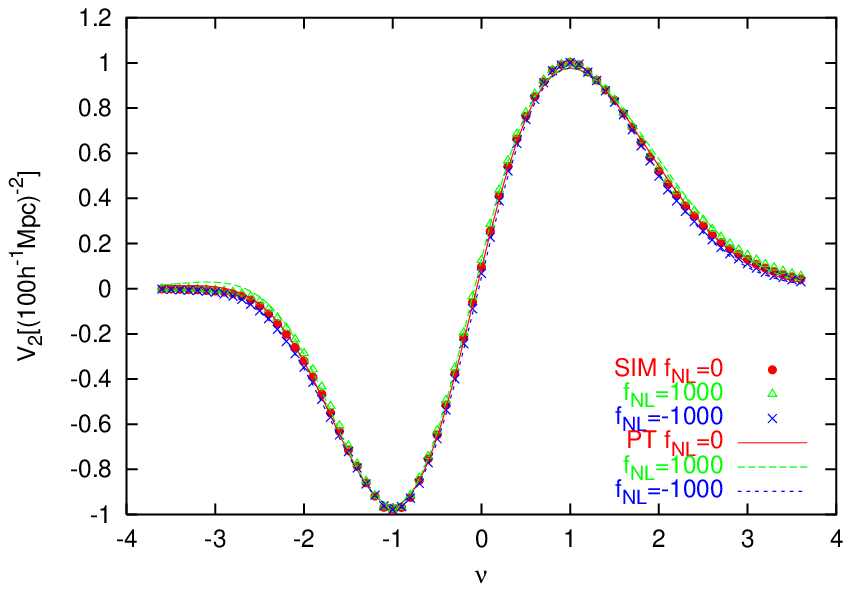}
\includegraphics[width=8cm]{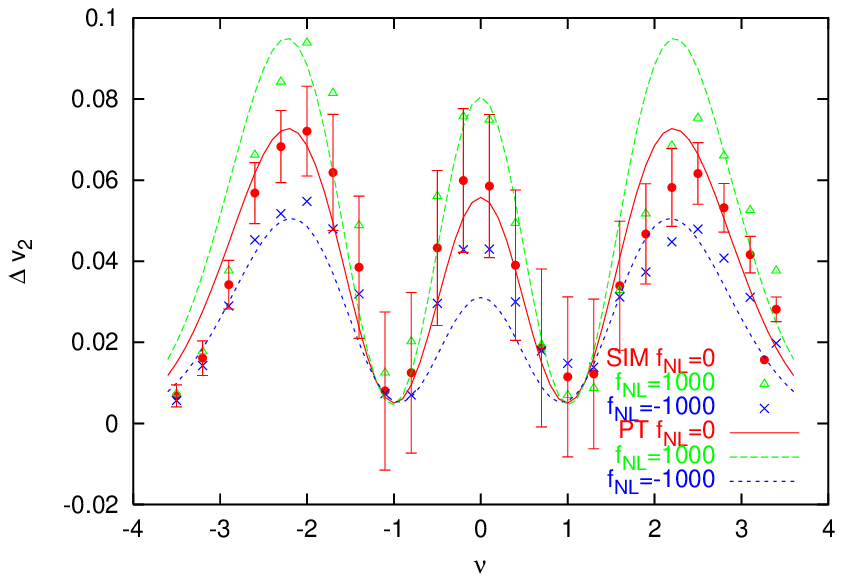}
\includegraphics[width=8cm]{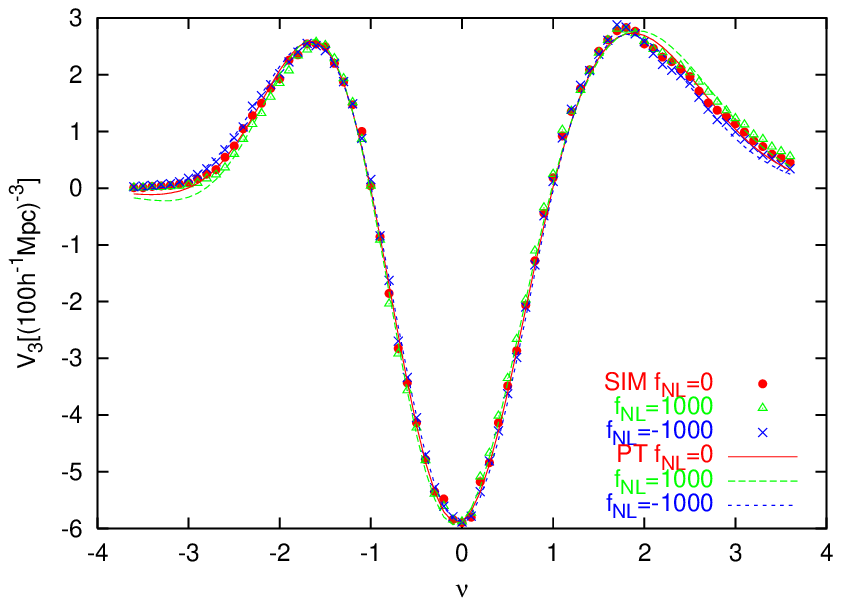}
\includegraphics[width=8cm]{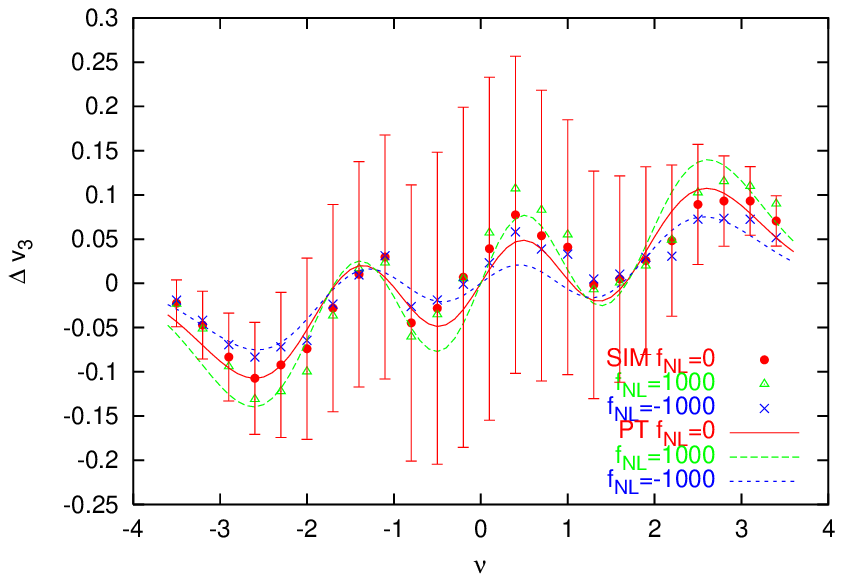}
\caption{Four Minkowski Functionals $V_k$ ({\it left}) and their
difference ratios $\Delta v_k$ ({\it right}) for the simulated mass
density fields at $z=3.96$, with $f_{\rm NL}=0$ (filled circles), 1000
(open triangles) and $-1000$ (crosses). The simulated fields are
smoothed with a Gaussian window function at the scale $R=$
10$h^{-1}$Mpc.  The error-bars denote the sample variance estimated
from 1000 Gaussian realizations with same $z, R$ and box-size as the
simulations.  For comparison, the theoretical expectations from
perturbation theory (equations [\ref{eq:MFs}] and
[\ref{eq:MFs_perturb}]) are plotted with lines.} \label{fig:2}
\end{center}
\end{figure*}
%%%%%%%%%%%%%%%%%%%%%%%%%%%%%%%%%%%%%%%%%%%%%%%%%%%%%%%

%%%%%%%%%%%%%%%%%%%%%%%%%%%%%%%%%%%%%%%%%%%%%%%%%%%%%%%
\begin{table*}
\centering
%\begin{minipage}{150mm}
\caption{Root-mean-square differences
between simulated MFs and perturbative formulae at different $z$ and
$f_{\rm NL}$ with the corresponding $\sigma_0$.  The smoothing scale
is fixed at $R=10h^{-1}$Mpc.}
\label{tab:dif}
\begin{tabular}{cccccccccccc}
\hline
& & & \multicolumn{4}{c}{$\sqrt{\langle(\Delta v_k^{(\rm SIM)}
-\Delta v_k^{(\rm PT)})^2\rangle}$} &
& \multicolumn{4}{c}{$\sqrt{\langle(\Delta v_k^{(\rm SIM)}-\Delta
v_k^{(\rm PT)})^2\rangle/\langle(\Delta v_k^{(\rm
SIM)})^2\rangle}$} \\
\cline{4-7}
\cline{9-12}
\raisebox{1.5ex}{$z$} &
\raisebox{1.5ex}{$f_{\rm NL}$} & \raisebox{1.5ex}{$\sigma_0$} &
\raisebox{-0.5ex}{$V_0$} & \raisebox{-0.5ex}{$V_1$} &
\raisebox{-0.5ex}{$V_2$} & \raisebox{-0.5ex}{$V_3$} & &
\raisebox{-0.5ex}{~~~$V_0$~~~} & \raisebox{-0.5ex}{~~~$V_1$~~~} &
\raisebox{-0.5ex}{~~~$V_2$~~~} & \raisebox{-0.5ex}{~~~$V_3$~~~} \\
\hline
5.15 & 0 & 0.080 & 0.003 & 0.006 & 0.009 & 0.024 & &
0.15 & 0.27 & 0.24 & 0.41 \\
5.15 & 1000 & 0.080 & 0.005 & 0.008 & 0.013 & 0.029 & &
0.17 & 0.27 & 0.24 & 0.38 \\
5.15 & $-1000$ & 0.080 & 0.003 & 0.006 & 0.012 & 0.024 & &
0.33 & 0.44 & 0.40 & 0.54 \\
3.96 & 0 & 0.099 & 0.003 & 0.006 & 0.009 & 0.023 & &
0.14 & 0.24 & 0.20 & 0.37 \\
3.96 & 1000 & 0.099 & 0.006 & 0.010 & 0.014 & 0.031 & &
0.20 & 0.34 & 0.26 & 0.41 \\
3.96 & $-1000$ & 0.099 & 0.003 & 0.004 & 0.008 & 0.020 & &
0.14 & 0.20 & 0.23 & 0.41 \\
2.13 & 0 & 0.16 & 0.007 & 0.014 & 0.021 & 0.044 & &
0.18 & 0.34 & 0.29 & 0.44 \\
2.13 & 1000 & 0.16 & 0.011 & 0.021 & 0.029 & 0.057 & &
0.23 & 0.40 & 0.34 & 0.48 \\
2.13 & $-1000$ & 0.16 & 0.004 & 0.010 & 0.017 & 0.034 & &
0.15 & 0.31 & 0.27 & 0.40 \\
0.96 & 0 & 0.24 & 0.015 & 0.030 & 0.043 & 0.081 & &
0.26 & 0.48 & 0.41 & 0.58 \\
0.96 & 1000 & 0.24 & 0.020 & 0.039 & 0.055 & 0.10 & &
0.31 & 0.54 & 0.48 & 0.65 \\
0.96 & $-1000$ & 0.24 & 0.010 & 0.023 & 0.034 & 0.065 & &
0.22 & 0.43 & 0.36 & 0.51 \\
0 & 0 & 0.38 & 0.035 & 0.067 & 0.095 & 0.17 & &
0.40 & 0.68 & 0.63 & 0.86 \\
0 & 1000 & 0.38 & 0.042 & 0.078 & 0.11 & 0.19 & &
0.44 & 0.73 & 0.69 & 0.93 \\
0 & $-1000$ & 0.38 & 0.028 & 0.058 & 0.083 & 0.15 & &
0.36 & 0.64 & 0.58 & 0.80 \\
\hline
\end{tabular}
%\end{minipage}
\end{table*}
%%%%%%%%%%%%%%%%%%%%%%%%%%%%%%%%%%%%%%%%%%%%%%%%%%%%%%%

Fig. \ref{fig:2} shows examples of MFs $V_k$ (left panels) and the
difference ratio $\Delta v_k$ (right panels) for simulated mass
distributions in the weakly non-linear regime.  We smooth on a scale
$R=10h^{-1}$Mpc which, at $z=3.96$, marks the transition to the
non-linear regime since the variance of the smoothed density
fluctuation $\sigma_0\simeq 0.1$. The different symbols show the
different $f_{\rm NL}$ of $0$ and $\pm 1000$. The error-bars
represent the sample variance estimated from 1000 Gaussian
realizations with the same $R$, $z$ and box-size as the simulations.
The perturbative formulae discussed above are plotted with lines for
comparison. Results for the simulations with $f_{\rm NL}=\pm 100$
and $\pm 500$ are found to be linearly scaled between those with
$f_{\rm NL}=0$ and $\pm 1000$.

The theoretical curves reproduce the features of the simulated MFs
very well.  We quantitatively estimate the agreement between the
simulation results $\Delta v_k^{\rm (SIM)}(\nu_i)$ and the
perturbative formulae $\Delta v_k^{\rm (PT)}(\nu_i)$ by calculating
the root-mean-square (rms) differences averaged over $i$.  Table
$\ref{tab:dif}$ lists the differences for each MF at different
redshifts $z$ (but $R$ is fixed to be $10h^{-1}$Mpc). The
differences are less than a few percent relative to the amplitude of
each MF (equation [\ref{eq:mfamp}]) when $\sigma_0<0.1$ and remains
at the $10$-percent level when $\sigma_0\sim 0.2$.  We also estimate
the rms differences divided by the rms of $\Delta v_k^{\rm
(SIM)}(\nu_i)$ averaged over $i$.  These quantities represent the
extent to which the theoretical predictions improve going from
linear theory to (2nd-order) perturbation theory.  The differences
between the 2nd-order perturbative predictions and the numerical
simulations is $0.15\sim 0.41$ times smaller than those
corresponding to linear theory at $\sigma_0 < 0.1$. These results
are consistent with the previous analysis by \citet{Nakagami2004}.

The differences between theory and simulations are quite small
compared to the sample variance. However, there is a systematic
feature, seen in the asymmetry of $V_0$ and $V_2$ with respect to
$\nu=0$; the perturbative predictions are symmetric. There are three
possible explanations for this effect. One is that higher-order
contributions - i.e. beyond 2nd-order - are significant.  Another
possibility arises from the use of the Zel'dovich approximation to
set the initial conditions of the simulations, which may be
responsible for an extra contribution to higher-order statistical
properties of clustering arising from transients \citep{Crocce2006}.
The other reason is the fact that the multivariate Edgeworth
expansion which is the basis of perturbation formulae has a limited
range of validity, especially at values of $\nu$ larger than unity
\citep{BK1995}. These effects must be considered carefully when
comparing with real survey results.

\subsection{Non-linear evolution and primordial non--Gaussianity}

%%%%%%%%%%%%%%%%%%%%%%%%%%%%%%%%%%%%%%%%%%%%%%%%%%%%%%%
\begin{figure*}
\begin{center}
\includegraphics[width=7cm]{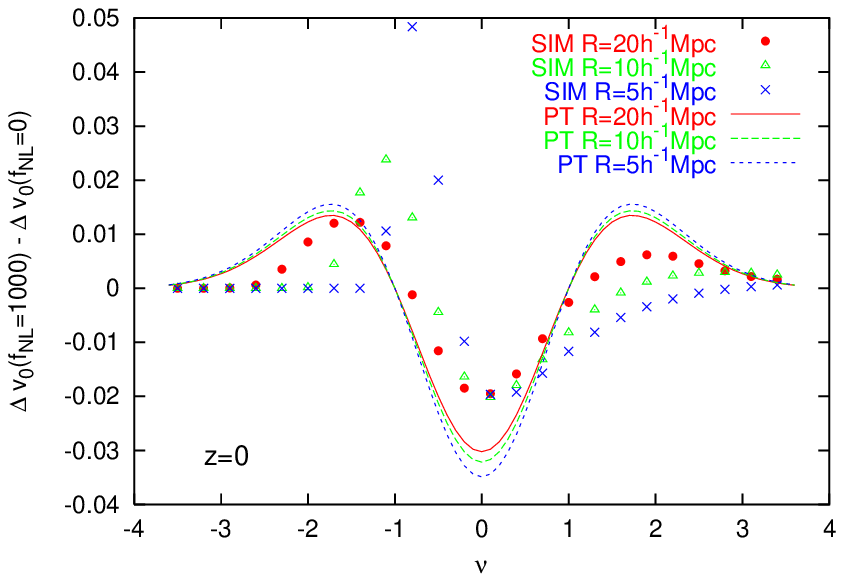}
\includegraphics[width=7cm]{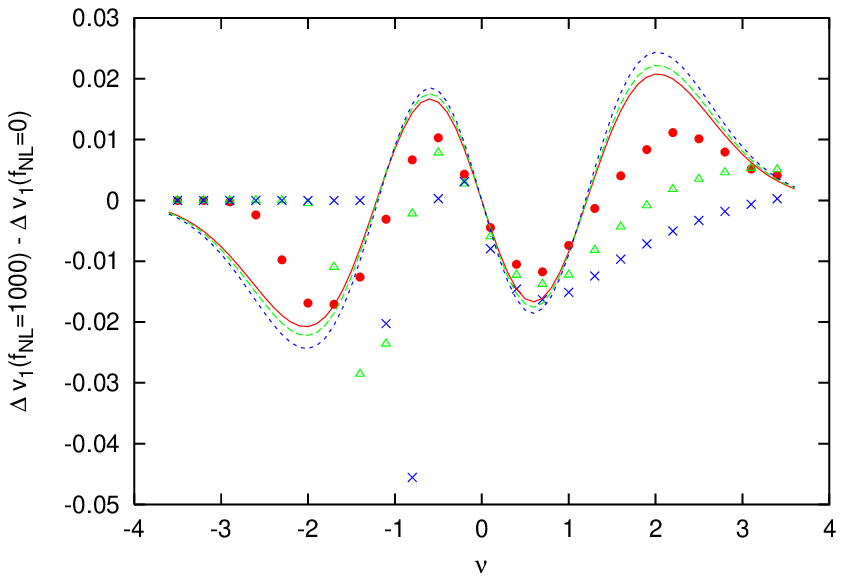}
\includegraphics[width=7cm]{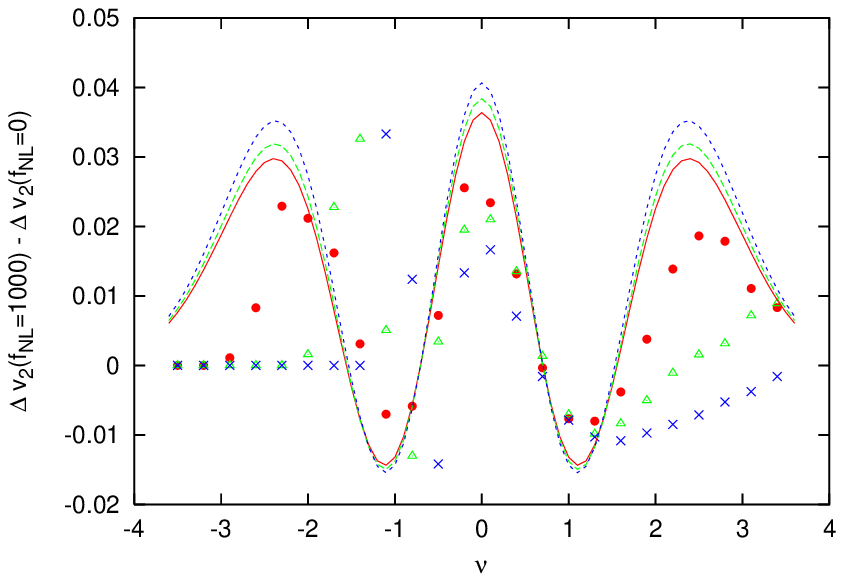}
\includegraphics[width=7cm]{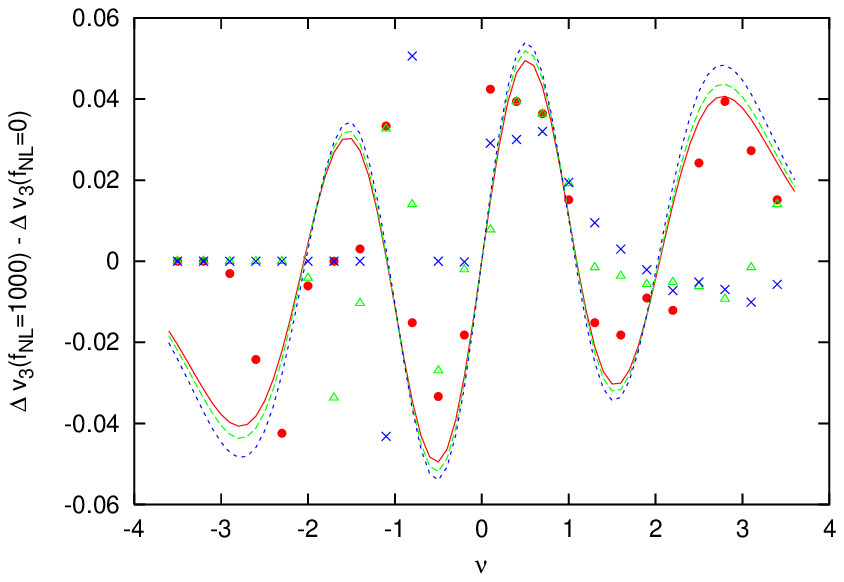}
\caption{The difference of $\Delta v_k$ with $f_{\rm NL}=1000$ from
those obtained with Gaussian initial conditions $\Delta v_k(f_{\rm
NL}=0)$ at $z=0$ for different smoothing scales $R=20h^{-1}$Mpc
$(\sigma_0=0.17)$, $10h^{-1}$Mpc $(\sigma_0=0.38)$, and $5h^{-1}$Mpc
$(\sigma_0=0.74)$. Simulated results averaged over three bins are
plotted with symbols and the perturbative formulae are also plotted
with lines.}
\label{fig:3}
\end{center}
\end{figure*}
%%%%%%%%%%%%%%%%%%%%%%%%%%%%%%%%%%%%%%%%%%%%%%%%%%%%%%%

In Fig. \ref{fig:3}, we focus on the differences between $\Delta
v_k$ with $f_{\rm NL}=1000$ and that with $f_{\rm NL}=0$ at $z=0$. The
perturbative predictions are also plotted for comparison.  The
deviation from the perturbative predictions becomes significant as the
smoothing scale $R$ is smaller ($\sigma_0$ increases) due to the
primordial non-Gaussian effect coupled with non-linear gravity.  The
increase of deviations at larger $\sigma_0$ are also seen quantitatively
in Table $\ref{tab:dif}$.  The shape of the deviation is skewed to the
positive side of $\nu$ with a higher peak at $\nu=-1/\sigma_0$ (the
number density is zero) while the overall amplitude of the deviation
$\Delta v_k$ is roughly the same as that from the perturbative
predictions.

It is interesting to estimate the sensitivity of the MFs to
primordial non-Gaussianity in the non-linear regime,  because the
effect of primordial non--Gaussianity on the MFs should become
increasingly significant as the sample variance decreases, i.e. at
smaller smoothing scales. The MFs are, however, strongly correlated
with each other among different bins of the threshold $\nu$ and it
is therefore necessary to take into account their covariance when
estimating the significance of the primordial non-Gaussian effect
with, e.g., chi-squared statistics. If the covariances among
different bins were not considered, one would overestimate the value
of chi-square as the total number of bins increases.  When the field
follows nearly Gaussian statistics, the covariance matrix is well
approximated with the one numerically estimated from a large number
of Gaussian realizations. \citep{Komatsu2003,HKM2006}.  When the
field is non-linearly evolved, it is an exceptionally time-consuming
process to generate enough number of realizations to compute the
inverse matrix of the covariance (the number of realizations must be
larger than the degree-of-freedom at least).

Instead of calculating the covariance matrix directly, therefore, we
instead estimate the amount of information contained in each MF as a
function of $\nu$.  For this purpose, we calculate the effective
number of bins $N_{\rm eff}$ for each MF and for all MFs combined as
follows:
%%%%%%%%%%%%%%%%%%%%%%%%%%%%%%%%%%%%%%%%%%%%%%%%%%%%%%%
\begin{equation}
\label{eq:neff}
N_{\rm eff}=N_{\rm bin}\frac{\sum_{i,j}^{N_{\rm bin}}\Delta v_i^{\rm
(PT)}(C^{-1})_{ij}\Delta v_j^{\rm (PT)}}{\sum_{i}^{N_{\rm bin}}
\Delta v_i^{\rm (PT)}C_{ii}^{-1}\Delta v_i^{\rm (PT)}}
\end{equation}
%%%%%%%%%%%%%%%%%%%%%%%%%%%%%%%%%%%%%%%%%%%%%%%%%%%%%%%
where $i$ and $j$ denote the binning number of different $\nu$ and
different kinds of MFs and $N_{\rm bin}$ denotes the total number of
bins.  The covariance matrix $C_{ij}=\langle\Delta v_i \Delta
v_j\rangle$ is computed from 1000 Gaussian realizations with the same
cosmological parameters and the same box-size as those of the $N$-body
simulations.  As $N_{\rm bin}$ is increased in a fixed range of $\nu$
from $-3.6$ to $3.6$, the values of $N_{\rm eff}$ converges to 2, 6,
8, and 12 for each MF from $k=0$ to 3 and then 12 for all MFs
combined.  The results indicate that the correlations among different
bins of $\nu$ is very strong for $V_0$ and that higher $k$-th MFs have
more independent information as a function of $\nu$.

Applying the value of $N_{\rm eff}$ for non-linearly evolved
simulations, we calculate the chi-square values of the primordial
non-Gaussian effect on MFs as a function of $f_{\rm NL}$ as
%%%%%%%%%%%%%%%%%%%%%%%%%%%%%%%%%%%%%%%%%%%%%%%%%%%%%%%
\begin{equation}
\label{eq:chisq}
\chi^2(f_{\rm NL})=\frac{N_{\rm eff}}{N_{\rm bin}}\sum_i^{N_{\rm bin}}
\frac{(\Delta v_i^{\rm (SIM)}(f_{\rm NL})
-\Delta v_i^{\rm (SIM)}(f_{\rm NL}=0))^2}{\langle\Delta
v_i^{\rm (SIM)}(f_{\rm NL}=0)^2\rangle}
\end{equation}
%%%%%%%%%%%%%%%%%%%%%%%%%%%%%%%%%%%%%%%%%%%%%%%%%%%%%%%
The variance $\langle \Delta v_i^{\rm (SIM)}(f_{\rm NL}=0)^2\rangle$
is estimated from $10$ realizations of $N$-body simulations with
Gaussian initial conditions (the cosmological parameters and
simulation box-size are the same as for the $N$-body simulations). The
normalized MFs $\Delta v_k(f_{\rm NL})$ at arbitrary $f_{\rm
NL}$ is linearly interpolated using the simulation results
with $f_{\rm NL}=0$ and $1000$.
We confirm that the linear interpolation works well using
simulations with $|f_{\rm NL}|=100$ and 500.

Table \ref{tab:fnl} lists the value of $f_{\rm NL}$ at different $R$
when the effect of the primordial non-Gaussianity is comparable to the
sample variance, that is $\chi^2=1$.  The volume of the simulation
box-size is $0.125(h^{-1}{\rm Gpc})^3$, which is less than half the
volume of the SDSS main galaxy sample $0.3(h^{-1}{\rm Gpc})^3$. As the
smoothing scale decreases, the primordial non-Gaussianity becomes
significant.  At $R=5h^{-1}$Mpc, the primordial non-Gaussianity with
$f_{\rm NL}=50$ is comparable to the sample variance and then
corresponds to the current observational constraints from WMAP. Note
that the detectability of primordial non--Gaussianity from actual
observations is, however, strongly dependent on the uncertainty of the
cosmological parameters and the galaxy biasing, which we have not
attempted to model in detail.

%%%%%%%%%%%%%%%%%%%%%%%%%%%%%%%%%%%%%%%%%%%%%%%%%%%%%%%
\begin{table}
\centering
\caption{The values of $f_{\rm NL}$ at $\chi^2=1$ when the
effect of the primordial non-Gaussianity is comparable to the sample
variance of mass density field for different smoothing scale $R$
(equation [\ref{eq:chisq}]). The values of $N_{\rm eff}$ for each MF
$V_k$ are 2, 6, 8, and 12 from $k=0$ to 3 and 12 for all MFs
combined. The volume of the density field is a cube at a length
$0.5h^{-1}$Gpc and the redshift is $0$.  The other cosmological
parameters are fixed to be fiducial values. The effective number of
bins $N_{\rm eff}$ (equation [\ref{eq:neff}]) is also listed in last
line.}
\label{tab:fnl}
\begin{tabular}{cccccc}
\hline
& \multicolumn{5}{c}{$f_{\rm NL}$ at $\chi^2=1$} \\
\cline{2-6}
\raisebox{1.5ex}{$R[h^{-1}$Mpc]}
& $V_0$ & $V_1$ & $V_2$ & $V_3$ & All MFs \\
\hline
30 & 770 & 480 & 520 & 370 & 350 \\
20 & 420 & 300 & 310 & 210 & 210 \\
10 & 190 & 180 & 140 & 150 & 110 \\
 5 &  90 &  80 &  90 &  60 &  50 \\
\hline
\end{tabular}
\end{table}
%%%%%%%%%%%%%%%%%%%%%%%%%%%%%%%%%%%%%%%%%%%%%%%%%%%%%%%

%%%%%%%%%%%%%%%%%%%%%%%%%%%%%%%%%%%%%%%%%%%%%%%%%%%%%%%
\section{Summary and Conclusions}
\label{sec:summary}

We have studied the imprint of primordial non--Gaussianity on the
topological properties of LSS using the MFs. Characterizing primordial
non--Gaussianity as a quadratic correction to the primordial potential
fluctuation with constant amplitude $f_{\rm NL}$, we compare the MFs
with different values of $f_{\rm NL}$ from the mildly to the strongly
non-linear regime using high-resolution $N$-body
simulations. Perturbative formulae of the MFs based on the
multivariate Edgeworth expansion well reproduce the MFs of simulated
mass density fields in the weakly non-linear regime. When the
amplitude of the density fluctuation $\sigma_0<0.1$ and $|f_{\rm
NL}|<1000$, the deviations of the perturbative formulae from
simulations are less than a few percent of the amplitude of each MF.
They are also $10\sim 40$ percent in respect to the non-Gaussian
contributions alone.

As the fluctuations become more strongly non-linear, the simulated
MFs begin to deviate significantly from the perturbative predictions
owing to non-linear gravitational evolution. In order to include
small-scale information in realistic cosmological data sets,
detailed numerical analysis is therefore essential.

When we include information from smaller scale fluctuations, the
effects of primordial non--Gaussianity are indeed significant. Using
$\chi^2$ statistics, we find that the primordial non-Gaussianity
with $f_{\rm NL}=50$ has  a significance level corresponding to
$1\sigma$, considering the sample variance of mass density fields at
$R=5h^{-1}$Mpc with a volume of $0.125(h^{-1}{\rm Gpc})^3$. This
implies that measuring the MFs in a SDSS-like survey could constrain
$f_{\rm NL}$ at a level comparable with current CMB limits. This is
an interesting result, since other observations, like the cluster
abundance, that can effectively constrain $f_{\rm NL}$ at high
redshifts, become useless at $z=0$ when non--Gaussian features
generated by non-linear dynamics completely obliterate primordial
ones \citep{Grossi2007,Kang2007}.

The actual detectability of the primordial non--Gaussianity is,
however, strongly dependent on the degeneracy between the cosmological
parameters and the primordial non--Gaussian effect. Understanding the
properties of the galaxy biasing is also very important in determining
the primordial non--Gaussianity accurately. We will consider this
issue in a forthcoming paper.

\section*{Acknowledgments}

We thank the anonymous referee for helpful comments. We thank
Takahiko Matsubara for useful discussions. C.H. acknowledges support
from the Particle Physics and Astronomy Research Council grant
number PP/C501692/1. Computations have been performed on the IBM-SP5
at CINECA (Consorzio Interuniversitario del Nord-Est per il Calcolo
Automatico), Bologna, with CPU time assigned under an INAF-CINECA
grant and on the IBM-SP4 machine at the ``Rechenzentrum der
Max-Planck-Gesellschaft'' at the Max-Planck Institut f\"ur
Plasmaphysik with CPU time assigned to the MPA.  We acknowledge
financial contribution from contracts ASI-INAF I/023/05/0, ASI-INAF
I/088/06/0 and INFN PD51.

%%%%%%%%%%%%%%%%%%%%%%%%%%%%%%%%%%%%%%%%%%%%%%%%%%%%%%%

\end{document}